\documentclass[english]{article}
\usepackage[T1]{fontenc}
\usepackage[latin1]{inputenc}
\usepackage{amsmath}
\usepackage{graphicx}
\usepackage{amssymb}

\makeatletter
\newcommand{\lyxaddress}[1]{
\par {\raggedright #1
\vspace{1.4em}
\noindent\par}
}

\usepackage{babel}
\makeatother
\begin{document}

\title{Magnetic field effects in the heat flow of charged fluids.}

\author{L. S. Garcia-Colín$^{1}$, A. Sandoval-Villalbazo$^{2,3}$, A. L.
García-Perciante$^{2}$ and A. Arrieta$^{3}$}

\maketitle

\lyxaddress{$^{1}$Departamento de Física, Universidad Autónoma Metropolitana-Iztapalapa,
Av. Purísima y Michoacán S/N, México D. F. 09340, México. Also at
El Colegio Nacional, Luis Gonzalez Obregón 23, Centro Histórico, México
D. F. 06020, México.}

\lyxaddress{$^{2}$Depto. de Matemáticas Aplicadas y Sistemas, Universidad Autónoma
Metropolitana-Cuajimalpa, Av. Pedro Antonio de los Santos No. 84,
México DF, México.}

\lyxaddress{$^{3}$Departamento de Física y Matemáticas, Universidad Iberoamericana,
Prolongación Paseo de la Reforma 880, México D. F. 01210, México.}

\date{\today{}}

\begin{abstract}
Heat conduction in ionized plasmas in the presence of magnetic fields
is today a fashionable problem. The kinetic theory of plasmas, in
the context of non-equilibrium thermodynamics, predicts a Hall-effect-like
heat flow due to the presence of a magnetic field in ionized gases.
This cross effect, the Righi-Leduc effect together with a heat flow
perpendicular to the magnetic field are shown to yield significant
contributions, under certain conditions to the ordinary Fourier component
of the heat flow. The thermal conductivities associated with these
effects change with the strength of the magnetic field for a given
temperature and density and are shown to be significant compared with
the parallel conductivity for a whole range of values of $\vec{B}$. 
\end{abstract}

\section{Introduction}

The behavior of charged particles in the presence of external magnetic
fields is a well known subject. According to the tenets of classical
non-equilibrium thermodynamics for not too large gradients, the simultaneous
heat and electrical flows in a magnetic field are linear functions
of the temperature and electrical potential gradients, respectively.
The coefficients appearing in these relations are, in general, second
rank tensors which are functions of the external field and consist
of a symmetrical part and an antisymmetric Al part \cite{key-1,key-2,key-3}.
The latter ones are called the {}``Hall vector'' in the case of
electrical conduction and the {}``Righi-Leduc'' vector in the case
of heat conduction. The former is a well known effect whereas the
latter one isn't. It was discovered in 1887 and first measured and
confirmed by Waldemar Voigt in 1903 (see Refs. \cite{key-2,key-3}).
Curiously enough it has been, ever since, hardly mentioned in the
literature. Not even the authors of a rather beautiful and striking
experiment published recently \cite{key-4} recognize that what they
have really detected is the Righi-Leduc effect.

In this paper we address ourselves to a rather different calculation
of the heat conduction effects in magnetized plasmas from the theoretical
point of view. Indeed, in the case of a dilute ionized gas in the
presence of weak magnetic fields for densities $n$ in the interval
$10^{3}\leq n\leq10^{8}\, m^{-3}$ and temperatures in the range $10^{3}<T<10^{7}\, K$
the magnitude of heat, charge and mass currents arising from these
and other so-called {}``cross effects'' may contribute by factors
which are not necessarily negligible when compared with those arising
from ordinary heat, mass and electrical conduction. This result seems
to differ from those obtained by Braginski \cite{key-5} while qualitatively
agree with those obtained by Balescu in his excellent and exhaustive
treatment on plasma transport processes \cite{key-6}. We shall come
back to this point a little later in this paper. Moreover, we believe
that the approach here offered to the general subject of collisional
transport processes in plasmas may provide a better understanding
of this subject at temperatures and densities where collisions are
favored in a non-relativistic framework.

The structure of the paper is as follows. In section II we shall briefly
review the kinetic concepts used throughout the work, section III
contains the bulk of the main calculations leading to the heat conduction
transport coefficients and in section IV the results are given together
with some concluding remarks.

\section{Kinetic model}

The kinetic model we use in our approach to the problem is the well
known one based on the Boltzmann equation as originally proposed by
Chapman and Cowling \cite{key-7}. The substantial difference is that
for the case of ionized gases they never developed their method up
to the stage of placing it within the framework of linear irreversible
thermodynamics. Although they derive elementary expressions for the
Righi-Leduc and the Nernst-Ettingshausen effects (see Ref. \cite{key-2})
they never pursued any comparison of their formulae with experiments,
much less studied any possible applications. Here we would like to
focus ourselves on the former and related effects.

Thus, the system here considered is a binary mixture of electrically
charged particles with masses and charges $m_{i}$ and $e_{i}$ for
$i=a,b$. The density of the system is low enough such that the kinetic
description is valid. For simplicity we shall set the charge of the
ions $Z=1$. For such a system, the evolution of the distribution
functions of the molecules is given by the Boltzmann equation:\begin{equation}
\frac{df_{i}}{dt}=\frac{\partial f_{i}}{\partial t}+\vec{v_{i}}\cdot\frac{\partial f_{i}}{\partial\vec{r}}+\frac{e}{m_{i}}\left(\vec{E}+\vec{v_{i}}\times\vec{B}\right)\cdot\frac{\partial f_{i}}{\partial\vec{v_{i}}}=\sum_{i,j=a}^{b}J\left(f_{i}f_{j}\right)\label{1}\end{equation}
 where the subscript $i$ indicates the species and $J\left(f_{i}f_{j}\right)$
is the collisional term representing collisions between different
and same species. The Lorentz force on the left side of Eq. (\ref{1})
is treated here as an external force where the magnitude of the magnetic
field is small enough such that collisions dominate over cyclotron
motion. We must clarify that both the electric and magnetic fields
appearing in this force contain the self consistent fields produced
by the plasma and are governed by the Maxwell's equations (see Ref.
\cite{key-6}). This weak field approximation implies $\omega_{i}\tau\approx1$
where $\omega_{i}=eB/m_{i}$ is the Larmor frequency, that is, the
frequency of the circular orbits that the particles describe around
magnetic field lines.

Once Eq. (\ref{1}) is defined, the derivation of the conservation
equations as well as the proof of the H-theorem are logically required.
The former is a standard step widely discussed in the literature (\cite{key-6}-\cite{key-10}).
The second result becomes a little bit more tricky due to the presence
of a magnetic field but we will not bother with it since it is not
essential to this paper \cite{key-8}. We here proceed directly with
the solution of Eq. (\ref{1}) following the standard Hilbert-Chapman-Enskog
approximation. Since a local Maxwellian distribution function $f_{i}^{(0)}$
is clearly a solution to the homogeneous part of Eq. (\ref{1}) we
propose that the single particle distribution functions $f_{i}$ ($i=a,\, b$)
may be expanded around $f_{i}^{(0)}$ in a power series of Knudsen's
parameter $\epsilon$ which, as well known, is a measure of the magnitude
of the macroscopic gradients \cite{key-7,key-9,key-10}. Thus\begin{equation}
f_{i}=f_{i}^{(0)}\left[1+\epsilon\varphi_{i}^{(1)}+\mathcal{O}\left(\epsilon^{2}\right)\right]\label{2}\end{equation}
 In Eq. (\ref{2}) the functional equilibrium assumption is also invoked,
namely the time dependence of $f_{i}^{(0)}$ and $\varphi_{i}^{(n)}$
for all $n$ occurs only through the conserved densities. The particle
density is $n_{i}\left(\vec{r},t\right)$ ($i=a,\, b$), the barycentric
velocity $\vec{u}\left(\vec{r},t\right)$ and the local temperature
$T\left(\vec{r},t\right)$ is in this paper assumed to be the same
for ions and electrons \cite{key-11}. $T\left(\vec{r},t\right)$
is related to the internal energy density $\varepsilon\left(\vec{r},t\right)$
by the standard ideal gas relationship. In this work we shall deal
only with the first order in the gradients correction to $f_{i}$
characterized by $\varphi_{i}^{(1)}$, namely the Navier-Stokes-Fourier
regime.

Substitution of Eq. (\ref{2}) into Eq. (\ref{1}), leads to order
zero in $\epsilon$ to the Euler equations of magnetohydrodynamics.
To first order in $\epsilon$ one obtains a set of two linear integral
equations for $\varphi_{i}^{(1)}$ which involve the linearized collision
kernels in their homogeneous terms whereas the inhomogeneous ones
contain a combination of terms involving the macroscopic gradients
$\nabla T$, $\nabla\vec{u}$ and the diffusive force $\vec{d_{ij}}=-\vec{d_{ji}}$.
Indeed, after a somewhat lengthy but standard manipulation \cite{key-7}
one gets that\begin{align}
\frac{m_{i}}{kT}\vec{c}_{i}^{^{^{\,\,0}}}\vec{c}_{i}:\nabla\vec{u}+\left[\left(\frac{m_{i}c_{i}^{2}}{2kT}-\frac{5}{2}\right)\frac{\nabla T}{T}+\frac{n_{i}}{n}\vec{d}_{ij}\right]\cdot\vec{c}_{i} & =-\frac{m_{i}}{\rho kT}\sum_{j=a}^{b}e_{j}\int d\vec{c}_{j}f_{j}^{(0)}\varphi_{j}^{(1)}\left(\vec{c}_{j}\times\vec{B}\right)\cdot\vec{c}_{i}\nonumber \\
 & \!\!\!\!\!\!\!\!\!\!\!\!\!\!\!\!\!\!\!\!\!\!\!\!\!\!\!\!\!\!\!\!\!\!\!\!\!\!\!\!\!\!\!\!\!\!-\frac{e_{i}}{m_{i}}\vec{c}_{i}\times\vec{B}\frac{\partial\varphi_{i}^{(1)}}{\partial\vec{v}_{i}}+\mathrm{C}\left(\varphi_{i}^{(1)}\right)+\mathrm{C}\left(\varphi_{i}^{(1)}\varphi_{j}^{(1)}\right)\qquad i=a,\, b\label{2.1}\end{align}
 where a superscript {}``$o$'' over a tensor indicates its symmetric
traceless part. In Eq. (\ref{2.1}), $\mathrm{C}\left(\varphi_{i}^{(1)}\right)$
and $\mathrm{C}\left(\varphi_{i}^{(1)}\varphi_{j}^{(1)}\right)$ are
the linearized collision kernels whose explicit forms are also well
known \cite{key-6}-\cite{key-10} and $\vec{d}_{ij}$ is the diffusive
vector force given by\begin{equation}
\vec{d}_{ab}=\nabla\frac{n_{a}}{n}+\frac{n_{a}n_{b}\left(m_{a}-m_{b}\right)}{n\rho}\frac{\nabla p}{p}-\frac{n_{a}n_{b}}{\rho\, p}\left(m_{b}e_{a}-m_{a}e_{b}\right)\cdot\vec{E'}\label{4}\end{equation}
 which satisfies the property $\vec{d}_{ij}=-\vec{d}_{ji}$. $\vec{E}'=\vec{E}+\vec{u}\times\vec{B}$
is the {}``effective'' electric force, $\vec{u}$ the barycentric
force defined as\begin{equation}
\rho\vec{u}\left(\vec{r},t\right)=\sum_{i=a}^{b}\rho_{i}\vec{u}_{i}\left(\vec{r},t\right)\label{5}\end{equation}
 $\rho_{i}=m_{i}n_{i}$ ($i=a,\, b$) and $\rho=\rho_{a}+\rho_{b}$.
Also $\vec{u}_{i}\left(\vec{r},t\right)=\left\langle \vec{v}_{i}\right\rangle $
where $\left\langle \right\rangle $ is the standard average taken
with $f_{i}\left(\vec{r},\vec{v}_{i},t\right)$.

We must emphasize that, as has been recently shown \cite{key-10},
it is only in this representation that the Onsager reciprocity relations
hold true, at least when $\vec{B}=0$. If $\vec{B}$ is different
from zero the proof of such relations starting from the linearized
integral equations for $\varphi_{i}^{(1)}$ will be discussed elsewhere.
The solution to Eqs. (\ref{2.1}) has been carefully outlined in Ref.
\cite{key-13}. Using Curie's theorem which allows setting $\nabla\vec{u}=0$,
the solutions are found to be of the form

\begin{equation}
\varphi_{i}^{(1)}=-\vec{\mathbb{A}}_{j}\cdot\frac{\nabla T}{T}-\vec{\mathbb{D}}_{i}\cdot{\bf d}_{ij}\qquad i=a,\, b\label{6.1}\end{equation}
 where\begin{equation}
\vec{\mathbb{A}}_{i}=\mathrm{A}_{i}^{(1)}\vec{c}_{i}+\mathrm{A}_{i}^{(2)}\left(\vec{c}_{i}\times\vec{B}\right)+\mathrm{A}_{i}^{(3)}\vec{B}\left(\vec{c}_{i}\cdot\vec{B}\right)\label{7.1}\end{equation}
 and a similar expression for $\vec{\mathbb{D}}_{i}$. The scalars
$\mathrm{A}_{i}^{(j)}$ ($j=1,\,2,\,3$) appearing in Eq. (\ref{7.1})
are functions of of $n,\, T,\, c_{i}^{2},\, B^{2}$ and $\left(\vec{c}_{i}\cdot\vec{B}\right)^{2}$.

When Eq. (\ref{6.1}) and its analog for $\vec{\mathbb{D}}_{i}$.
are substituted in Eq. (\ref{2.1}) we get a set of linear integral
equations for the unknown functions $\mathrm{A}_{i}^{(j)}$ and their
analogs for the $\vec{\mathbb{D}}_{i}$ vector. As shown in Ref. \cite{key-13},
these equations are\begin{equation}
f_{i}^{(0)}\left(\frac{m_{i}c_{i}^{2}}{2kT}-\frac{5}{2}\right)\vec{c}_{i}=f_{i}^{(0)}\left\{ \mathrm{C}\left(\vec{c}_{i}\mathcal{R}_{i}\right)+\mathrm{C}\left(\vec{c}_{i}\mathcal{R}_{i}+\vec{c}_{j}\mathcal{R}_{j}\right)\right\} \label{8.1}\end{equation}
 where $\mathcal{R}_{i}=\mathrm{A}_{i}^{(1)}+B^{2}\mathrm{A}_{i}^{(3)}$,
$i=a,\, b$ and\begin{equation}
f_{i}^{(0)}\left(\frac{m_{i}c_{i}^{2}}{2kT}-\frac{5}{2}\right)\vec{c}_{i}=-f_{i}^{(0)}\frac{m_{i}}{\rho kT}iB\vec{c_{i}}\mathcal{G}-f_{i}^{(0)}\frac{e_{i}}{m_{i}}\vec{c}_{i}B\mathcal{A}_{i}+f_{i}^{(0)}\left\{ \mathrm{C}\left(\vec{c}_{i}\mathcal{A}_{i}\right)+\mathrm{C}\left(\vec{c}_{i}\mathcal{A}_{i}+\vec{c}_{j}\mathcal{A}_{j}\right)\right\} \label{9.1}\end{equation}
 In Eq. (\ref{9.1})\begin{equation}
\mathcal{A}_{i}=\mathrm{A}_{i}^{(1)}+iB\mathrm{A}_{i}^{(2)}\label{10.1}\end{equation}
 \begin{equation}
\mathcal{G}=\mathrm{G}_{B}^{(1)}+iB\mathrm{G}_{B}^{(2)}\label{11.1}\end{equation}
 and\begin{equation}
\mathrm{G}_{B}^{(k)}=\frac{1}{2}\sum_{j=a}^{b}e_{j}\int d\vec{c}_{j}f_{j}^{(0)}\mathrm{A}_{j}^{(k)}\left[c_{j}^{2}-\frac{1}{B^{2}}\left(\vec{c}_{j}\cdot\vec{B}\right)^{2}\right]\label{12.1}\end{equation}
 for $k=1,\,2$.

Similar results are obtained for the unknown functions $\mathrm{D}_{i}^{(j)}$
appearing in the analog of Eq. (\ref{7.1}) for the vector $\vec{\mathbb{D}}_{i}$
but we shall not bother with them here since we will concentrate only
in heat conduction in the plasma. The $\vec{\mathbb{D}}_{i}$vector
is clearly related to diffusion phenomena \cite{key-8}. Equations
(\ref{8.1}) and (\ref{9.1}) are the basic ingredients required to
discuss the problem of heat conduction. The curious reader will immediately
realize the difference between these basic results and those used
by Braginski in his paper on this subject {[}Ref. \cite{key-5} pages
245-247]. In our method the unknown $\mathrm{A}_{i}^{(j)}$ functions
satisfy integral equations, (\ref{8.1}) and (\ref{9.1}), where the
collisional dynamics obeyed by the particles of the same and different
species are contained in the linearized collision kernels about which
no assumptions have yet been introduced. We shall now proceed in the
next section to study the heat conduction in the dilute plasma.

\section{Heat conduction in a fully ionized plasma}

According to classical irreversible thermodynamics \cite{key-1,key-3,key-14}
the expression for the heat flux vector in a multicomponent system
is given by\[
{\vec{J}'}_{q}=\vec{J}_{q}-\sum_{i}h_{i}\frac{\vec{J}_{i}}{m_{i}}\]
 where $\vec{J}_{i}$ is the diffusion vector for species $i$ and
$h_{i}$ its enthalpy. Since for an ideal gas $h_{i}=\frac{5}{2}kT$,
where $k$ is the Boltzmann constant, using the standard definition
for $\vec{J}_{q}$, namely for our case\[
\vec{J}_{q}=\sum_{i=a}^{b}\frac{m_{i}}{2}\left\langle c_{i}^{2}\right\rangle \]
 we get that\begin{equation}
\frac{1}{kT}{\vec{J}'}_{q}=\sum_{i=a}^{b}\int\left(\frac{m_{i}c_{i}^{2}}{2kT}-\frac{5}{2}\right)f_{i}\vec{c}_{i}d\vec{c_{i}}\label{13.1}\end{equation}
 since $\vec{J}_{i}=m_{i}\left\langle \vec{c}_{i}\right\rangle $.
Also, ${\vec{J}'}_{q}=0$ for $f_{i}=f_{i}^{(0)}$ so that, substituting
Eq. (\ref{6.1}) in Eq. (\ref{13.1}) and ignoring diffusive contributions
to ${\vec{J}'}_{q}$ we get that\begin{equation}
\left(kT\right)^{-1}{\vec{J}'}_{q}=\sum_{i=a}^{b}\left(\frac{m_{i}c_{i}^{2}}{2kT}-\frac{5}{2}\right)f_{i}^{(0)}\varphi_{i}^{(1)}\vec{c}_{i}d\vec{c_{i}}\label{14.1}\end{equation}
 where $\varphi_{i}^{(1)}$ is now given by Eqs. (\ref{6.1}) and
(\ref{7.1}).

To continue with the calculation of ${\vec{J}'}_{q}$ we now resort
to the almost orthodox method in kinetic theory, namely to expand
the unknown functions $\mathrm{A}_{j}^{(k)}$ in terms of a complete
set of orthonormal functions, the Sonine polynomials, so that

\begin{equation}
\mathrm{A}_{j}^{(k)}=\sum_{m=0}^{\infty}a_{j}^{(k)(m)}S_{3/2}^{(m)}\left(c_{i}^{2}\right)\qquad k=1,\,2,\,3\label{15.1}\end{equation}
 where the coefficients $a_{j}^{(k)(m)}$ are still to be determined
from the integral equations (\ref{8.1}) and (\ref{9.1}) and thus
depend on the interaction (Coulomb) potential between the species.
Notice however that in spite of the complicated form for $\vec{\mathbb{A}}_{i}$
given in Eq. (\ref{7.1}), when Eqs. (\ref{6.1}) and (\ref{7.1})
are introduced into Eq. (\ref{14.1}) all integrals over $\vec{c}_{i}$
have the same structure, namely\begin{equation}
\sum_{i=a}^{b}\left(\frac{m_{i}c_{i}^{2}}{2kT}-\frac{5}{2}\right)\frac{c_{i}^{2}}{3}f_{i}^{(0)}\sum_{m=0}^{\infty}a_{i}^{(k)(m)}S_{3/2}^{(m)}\left(c_{i}^{2}\right)d\vec{c_{i}}=-\frac{5}{2}kT\sum_{i=a}^{b}\frac{n_{i}}{m_{i}}a_{i}^{(k)(m)}\qquad k=1,\,2,\,3\label{16.1}\end{equation}
 where use has been made of the well known property that \cite{key-6}-\cite{key-10}\begin{equation}
\int_{0}^{\infty}e^{-x^{2}}S_{n}^{(p)}\left(x\right)S_{n}^{(q)}\left(x\right)x^{2n+1}dx=\frac{\Gamma\left(n+p+q\right)}{2p!}\delta_{pq}\label{17.1}\end{equation}
 and that $S_{n}^{(0)}\left(x\right)=1$. Clearly then\begin{equation}
{\vec{J}'}_{q}=-\frac{5}{2}k^{2}T\sum_{i=a}^{b}\frac{n_{i}}{m_{i}}\left[a_{i}^{(1)(1)}\nabla T+a_{i}^{(2)(1)}\vec{B}\times\nabla T+a_{i}^{(3)(1)}\vec{B}\left(\vec{B}\cdot\nabla T\right)\right]\label{18.1}\end{equation}
 Equation (\ref{18.1}) is an important result. Of the infinite number
of coefficients required to specify the functions $\mathrm{A}_{i}^{(k)}$
appearing in Eq. (\ref{15.1}) only one is required to compute the
explicit form of the constitutive equation (\ref{18.1}). This fact
readily simplifies the solutions to the integral equations (\ref{8.1})
and (\ref{9.1}) as we shall see below. However, before doing so let
us examine Eq. (\ref{18.1}). If we take $\vec{B}$ in the direction
of the $z$-axis, $\vec{B}\times\nabla T$ is a vector perpendicular
to both $\vec{B}$ and $\nabla T$ whereas $\vec{B}\left(\vec{B}\cdot\nabla T\right)=B^{2}\frac{\partial T}{\partial z}\vec{k}$
so that we may write that\begin{equation}
{\vec{J}'}_{q}=-\frac{5}{2}k^{2}T\sum_{i=a}^{b}\frac{n_{i}}{m_{i}}\left[\left(a_{i}^{(1)(1)}+B^{2}a_{i}^{(3)(1)}\right)\nabla_{\parallel}T+a_{i}^{(1)(1)}\nabla_{\perp}T+Ba_{i}^{(2)(1)}\nabla_{s}T\right]\label{19.1}\end{equation}
 where the last term represents a heat flow in the direction perpendicular
to both $\vec{B}$ and $\nabla T$. This is the well known Righi-Leduc
effect \cite{key-1}-\cite{key-3}. Notice also that when $\vec{B}=0$\begin{equation}
{\vec{J}'}_{q}=-\frac{5}{2}k^{2}T\sum_{i=a}^{b}\frac{n_{i}}{m_{i}}a_{i}^{(1)(1)}\nabla T=-\kappa_{\parallel}\nabla T\label{20.1}\end{equation}
 which is the well known form for Fourier's heat conduction equation
and the thermal conductivity is given by\begin{equation}
\left(\kappa_{\parallel}\right)_{B=0}=\frac{5}{2}k^{2}T\sum_{i=a}^{b}\frac{n_{i}}{m_{i}}a_{i}^{(1)(1)}\label{21.1}\end{equation}

In Eq. (\ref{21.1}) the coefficients $a_{a}^{(1)(1)}$ and $a_{b}^{(1)(1)}$
must arise from the solution to the integral equations (\ref{8.1})
and (\ref{9.1}) which indeed become identical to each other when
$\vec{B}=0$. In its more general form, Eq. (\ref{19.1}) thus reads
as\begin{equation}
{\vec{J}'}_{q}=-\kappa_{\parallel}\nabla_{\parallel}T-\kappa_{\perp}\nabla_{\perp}T-\kappa_{s}\nabla_{s}T\label{22.1}\end{equation}
 where the three conductivities are readily identified from Eq. (\ref{19.1}).
Once more a word of caution. Equation (\ref{22.1}) appears to be
the same as the one quoted by Braginski {[}see Eq. (4.33) of Ref.\cite{key-5}]
however, the evaluation of the coefficients $a_{i}^{(1)(m)}$ in our
case radically differs from the procedure followed by this author.
And worst, his definition of heat flux is foreign to the one used
here, that is, not in agreement with irreversible thermodynamics.

To determine the $a_{i}^{(1)(m)}$ coefficients we need to solve Eqs.
(\ref{8.1}) and (\ref{9.1}). The former one is straightforward and
has been discussed in the literature, specially in Appendix B of Ref.
\cite{key-13}. The second one, however poses some problems due to
the structure of its inhomogeneous term. Nevertheless by a subtle
generalization of the procedure followed to solve Eq. (\ref{8.1}),
Eq. (\ref{9.1}) is also solved using a variational procedure apparently
due to Davison \cite{key-15}\cite{key-16}. The steps are also outlined
in the same Appendix B of Ref. \cite{key-13}. We shall only quote
the results here,\[
a_{a}^{(1)(0)}+B^{2}a_{a}^{(3)(0)}=2.94\tau\]
 \begin{equation}
a_{a}^{(1)(1)}+B^{2}a_{a}^{(3)(1)}=1.96\tau\label{23.1}\end{equation}
 \[
a_{b}^{(1)(1)}+B^{2}a_{b}^{(3)(1)}=\frac{0.058}{M_{1}}\tau\]
 also calling $a_{i}^{(m)}\equiv a_{i}^{(1)(m)}+iBa_{i}^{(2)(m)}$,
$i=a,\, b$, $m=1,\,2$ following Eqs. (\ref{10.1}) and (\ref{15.1})
we get, after separating real and imaginary parts that\[
\mathrm{Re}\left[a_{a}^{(0)}\right]=a_{a}^{(1)(0)}=\left(56.3-662x^{2}-2.25x^{4}\right)\frac{\tau}{\Delta_{1}}\]
 \[
\mathrm{Im}\left[a_{a}^{(0)}\right]=Ba_{a}^{(2)(0)}=\left(647x+2.2x^{3}\right)\frac{\tau}{\Delta_{1}}\]
 \begin{equation}
\mathrm{Re}\left[a_{a}^{(1)}\right]=a_{a}^{(1)(1)}=\left(37.5+2147x^{2}+7.3x^{4}\right)\frac{\tau}{\Delta_{1}}\label{24.1}\end{equation}
 \[
\mathrm{Im}\left[a_{a}^{(1)}\right]=Ba_{a}^{(1)(2)}=\left(206x+2649x^{3}+9x^{5}\right)\frac{\tau}{\Delta_{1}}\]
 \[
\mathrm{Re}\left[a_{b}^{(1)}\right]=a_{b}^{(1)(1)}=\left(1.12+121.2x^{2}+154.4x^{4}\right)\frac{\tau}{M_{1}\Delta_{1}}\]
 \[
\mathrm{Im}\left[a_{b}^{(1)}\right]=Ba_{b}^{(1)(2)}=-\left(0.06x+7x^{3}+9x^{5}\right)\frac{\tau}{M_{1}\Delta_{1}}\]
 where \begin{equation}
\Delta_{1}=19+2078x^{2}+2650x^{4}+9x^{6}\label{25.1}\end{equation}
 and $x=\omega_{e}\tau=1.76\times10^{11}B\tau$ where $B$ is given
in teslas. $\tau$ is the mean free time obtained from the only independent
collision integral and is defined as%
\footnote{The numerical factors appearing in the six polynomials quoted in Eqs.
(\ref{24.1}) substitute those published in Appendix B of Ref. \cite{key-13}
which result from a revised calculation.%
} \begin{equation}
\tau=\frac{4\left(2\pi\right)^{3/2}\sqrt{m_{e}}\left(kT\right)^{3/2}\epsilon_{0}^{2}}{ne^{4}\psi}\label{26.1}\end{equation}
 In Eq. (\ref{26.1}) $\psi$ is the so called logarithmic function
which arises from using Debye's length as a cutoff length in the collision
integrals, defined as\begin{equation}
\psi=\ln\left[1+\left(\frac{16\pi kT\lambda_{D}\epsilon_{0}}{e^{2}}\right)^{2}\right]\label{27.1}\end{equation}
 where $\lambda_{D}$ is the Debye's length, namely\begin{equation}
\lambda_{D}=\sqrt{\frac{\epsilon_{0}kT}{ne^{2}}}\label{28.1}\end{equation}
 Notice that the consistency requirement that $\left(a_{a}^{(1)(1)}\right)_{B=0}=\left(\mathrm{Re}\left[a_{a}^{(1)}\right]\right)_{B=0}$
is satisfied within the approximation used here.

\noindent Therefore, summarizing the three thermal conductivities
in Eq. (\ref{22.1}) are given by\begin{equation}
\kappa_{\parallel}=\frac{5}{4}\frac{k^{2}T}{m_{e}}\times2.01n\tau\label{29.1}\end{equation}
 \begin{equation}
\kappa_{\perp}=\frac{5}{4}\frac{k^{2}T}{m_{e}}\frac{n\tau}{\Delta_{1}}\times\left(38.7+2270x^{2}+161x^{4}\right)\label{30.1}\end{equation}
 \begin{equation}
\kappa_{s}=\frac{5}{4}\frac{k^{2}T}{m_{e}}\frac{n\tau}{\Delta_{1}}\times\left(206x+2644x^{3}\right)\label{31.1}\end{equation}
where $\psi$ is defined in Eq. (\ref{27.1}).

We emphasize that Eqs. (\ref{29.1})-(\ref{31.1}) are valid for a
fully ionized hydrogen plasma ($n_{a}=n_{b}=n/2$, $m_{b}\gg m_{a}=m_{e}$)
in a first order in the gradients approximation, namely the Navier-Stokes-Fourier
regime. These are the main results of this paper.

\section{Discussion of the results}

The first thought that a reader may have concerning the nature of
Eqs. (\ref{29.1})-(\ref{31.1}) is to ask how they compare to those
obtained from the well accepted and dominant calculations obtained
much earlier by Spitzer \cite{key-17} and by Braginski \cite{key-5}.
As already pointed out in full detail by Balescu in Ref. \cite{key-6}
and emphasized by us in a recent paper \cite{key-13} this is quite
difficult. Neither of both authors, besides using the Fokker-Planck
and Landau kinetic equations, respectively, performed their calculations
within the framework of classical irreversible thermodynamics. In
particular, the correct definition for the heat flux in a multicomponent
mixture, e. g. Eq. (\ref{12.1}), was ignored. Therefore the expressions
that they quote for the thermal conductivities are not equivalent
to ours.

The most interesting feature of our results is that the three thermal
conductivities, exhibit a behavior which is similar to that shown
by the results obtained by Balescu who used a Landau type kinetic
equation which he solved using the {}``moments'' method. This result
is gratifying. It shows indeed that using the full Boltzmann equation
and solving the ensuing integral equations which define the coefficients
appearing in the explicit results for the transport coefficients leads
to results that approximately equal to those obtained from the Landau
equation when the practically {}``exact'' 21 moment approximation
is used. Comparing figure 1 of our paper with figure 5.1 in Balescu's
book confirms this statement. Thus, his prediction that different
methods used to compute transport coefficients will yield results
that will be within a 10\% difference from each other turns out to
be sustained.

Another issue is important. Working with the moment method leads to
results which are not clearly related to the order of the macroscopic
gradients in the system. This was first pointed out by Grad in his
monumental work on this subject \cite{key-23}. Referring to the quantity
that appears in his case namely, the one resulting from the collision
integral in the thirteen moment solution to the Boltzmann equation,
he asserts that {}``such a parameter has a number of interpretations''
(see p. 271-72 of Ref. \cite{key-23}) . Further, if one wants to
classify the resulting equations in terms of power in the gradients
one must apply the Chapman-Enskog expansion not to the Boltzmann equation
but to the moment equations. It is only then, as has been extensively
discussed in Ref. \cite{key-19} one may extract the Euler, Navier-Stokes-Fourier,
Burnett and higher order in the gradients contributions. Therefore
Balescu's nearly exact results as he claims, obtained with the 21
moment method, are not clearly related to the ordinary hydrodynamic
hierarchy of equations. We believe this is the main reason why our
results are somewhat different from his. It is outside the scope of
this paper to attempt a detailed comparison of both methods, mainly
a purely algebraic issue.

Finally we wish to remark also that these results together with similar
ones here obtained for the Dufour coefficient may be useful in accounting
for dissipative phenomena which are becoming rather important in the
physics of the intracluster medium \cite{key-20}-\cite{key-22}.

\includegraphics{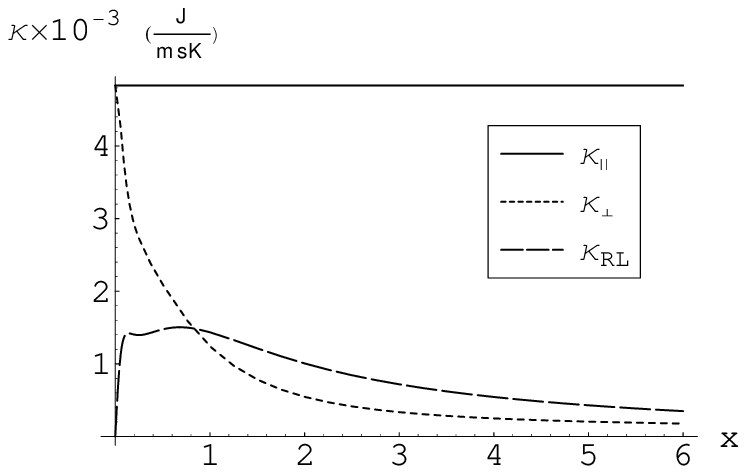}

{\small Fig. 1. The three thermal conductivities as functions of $x=\omega_{e}\tau$
for $T=10^{6}K$ and $n=10^{15}m^{-3}$.}{\small \par}


\begin{thebibliography}{10}
\bibitem{key-1}S. R. de Groot; {}``Thermodynamics of Irreversible
Process'' North Holland Publishing Co. Amsterdam 1952.

\bibitem{key-2}D. Miller; Chem. Rev. \textbf{60}, 15 (1960). This
lucid account on the validity of the Onsager relations has been reproduced
as Chap. 11 of the book {}``Transport Phenomena in Fluids'', H.
J. M. Hanley, ed M. Dekker, New York 1964.

\bibitem{key-3}H. Hasse; {}``Irreversible Thermodynamics'' Addison-Wesley
Publ. Co. Reading Mass. 1969.

\bibitem{key-4}I. Strohm, P. Wylder and G. Rikken; Phys. Rev. Lett.
\textbf{95}, 155901 (2005).

\bibitem{key-5}S. I. Braginski, {}``Transport Processes in a Plasma''
Plasma Physics Reviews (Consultants Bureau Enterprise, N. Y., 1965)
p. 205.

\bibitem{key-6}R. Balescu, {}``Transport Processes in Plasmas Vol.
I: Classical Transport'', North Holland Publ. Co. Amsterdam (1988).

\bibitem{key-7}S. Chapman and T. G. Cowling; {}``The Mathematical
Theory of Non-Uniform Gases'' Cambridge Univ. Press, Cambridge Third
Edition 1970.

\bibitem{key-8}L. S. García-Colín and L. Dagdug; {}``The Kinetic
Theory of a Dilute Ionized Plasma'', submitted to Cambridge University
Press (2007).

\bibitem{key-9}G. Uhlenbeck and G. W. Ford; {}``Lectures in Statistical
Mechanics''; Am. Math. Society, Providence, RI 1963 (Chap. VI).

\bibitem{key-10}J. R. Dorfman and H. van Beijeren; {}``The Kinetic
Theory of Gases in Statistical Mechanics''; Pt. B; Bruce Berne, ed.
Plenum Press, N. Y. 1977 pp. 65.

\bibitem{key-11}No difficulties would arise in assuming that the
ion temperature $T_{i}$ is not equal to the electron temperature
$T_{e}$, it would only complicate the algebra. For a discussion see
Ref. \cite{key-5}.

\bibitem{key-12}P. Goldstein and L. S. García-Colín, J. Non-Equilib.
Thermodyn. \textbf{30}, 173 (2005).

\bibitem{key-13}L. S. García-Colín, A. L. García-Perciante and A.
Sandoval-Villalbazo; Phys. Plasmas \textbf{14}, 012305 (2007).

\bibitem{key-14}J. Hirschfelder, C. F. Curtiss and R. B. Byrd; {}``The
Molecular Theory of Liquids and Gases''; J. Wiley and Sons, N. Y.
Second Ed. 1964.

\bibitem{key-15}B. B. Robinson and I. B. Bernstein; Ann. Phys. \textbf{18},
110 (1962).

\bibitem{key-16}P. Clemmow and J. P. Dougherty; {}``Electrodynamics
of Particles and Plasmas'' 2nd ed; Addison-Wesley, Reading Mass.
(1990).

\bibitem{key-17}L. Spitzer, Jr. {}``The Physics of Fully Ionized
Gases'' Wiley-Intescience Publ. Co. N. Y., 1962 and Ref. {[}9].

\bibitem{key-18}R. M. Velasco, F. J. Uribe and L. S. Garcia-Colin;
Phys. Rev. E \textbf{66}, 032103 (2002).

\bibitem{key-23}H. Grad, {}``Principles of the Kinetic Theory of
Gases'', Handbook of Physics S. Flügge, Ed., Springer Verlag, Berlin
(1958) Band XXII.

\bibitem{key-19}L. S. Garcia-Colin, R. M. Velasco and F. J. Uribe;
J. Non-Equilib. Thermodyn \textbf{29}, 257 (2004).

\bibitem{key-20}M. Brüggen and M. Ruszkowski; astro-ph/0512148 and
Astrophys. J. 615, 675 (2004).

\bibitem{key-21}A. C. Fabian, C. S. Reynolds, G. B. Taylor and R.
J. H. Dunn; Mon. Not. R. Astron. Soc. 363 177 (2005).

\bibitem{key-22}S. A. Balbus and J. F. Hawley, Rev. Mod. Phys. \textbf{70},
1 (1998). 
\end{thebibliography}
\end{document}